\documentclass[twocolumn,showpacs,aps,amsmath,amssymb]{revtex4}
\usepackage{graphicx}
\usepackage{dcolumn}
\usepackage{bm}
\usepackage{graphicx}
\usepackage{amsmath}
\begin{document}
\title{Aharonov-Bohm Effect and Disclinations in an Elastic Medium}
\author{ Claudio Furtado$^{1}$\footnote{Electronic Address: 
furtado@fisica.ufpb.br}, ~A.~M. de~M. Carvalho$^{2}$
\footnote{Electronic Address: ammc@uefs.br} 
and ~C. ~A. de Lima Ribeiro$^{2}$\footnote{Electronic Address: calr@uefs.br} }
\affiliation{$^{1}$Departamento de F\'{\i}sica, CCEN,  Universidade Federal
da Para\'{\i}ba, Cidade Universit\'{a}ria, 58051-970 Jo\~ao Pessoa, PB, Brazil\\
$^{2}$Departamento de F\'{\i}sica, Universidade Estadual de Feira de Santana, 
BR116-Norte, Km 3, 
44031-460, Feira de Santana, BA, Brazil}
\begin{abstract}
In this work we investigate quasiparticles in the background of defects in 
solids using the geometric 
theory of defects. We use  the parallel transport matrix to 
study the Aharonov-Bohm effect 
in this background. For quasiparticles moving in this effective medium we demonstrate 
an effect  similar to the gravitational Aharonov-
Bohm effect. We analyze this effect in 
an elastic medium with  one and $N$ defects.
\end{abstract}
\pacs{04.90.+e, 04.20.-q,02.40.-k  }
\maketitle
\section{Introduction}
The first approach to the theory of defects in an elastic media was done by the 
Italian school with the development of the theory of dislocations  in 1900. 
After those initial approaches, many theories and experiments were put forward 
in order to describe and to observe defects in solids~\cite{nabarro}. In recent 
years a series of articles, inspired by the pioneering ideas of Kr\"oner~\cite
{kroner} and Bilby {\it et al.}~\cite{bilby}, developed a geometric theory of 
defects. In this framework the elastic solids with topological defects can be 
described by Riemann-Cartan geometry~\cite{kleinert,anp:dia,katanaev92,6}.

In this formulation we use the techniques of differential geometry to describe 
the strain and stress induced by the defect in an elastic medium. All this 
information is contained in the geometric quantities (metric, curvature tensor, 
etc.) that describes the elastic medium with defects. The boundary conditions 
imposed by the defect, in the elastic continuum medium, are accounted for by a non-
Euclidean metric.  In the continuum limit, the solid can be viewed as a Riemann-
Cartan manifold. In general, the defect corresponds to a singular curvature or 
torsion (or both) along the  defect line~\cite{katanaev92}, where the curvature 
and the torsion of the manifold are associated to the topological defects, 
disclinations and dislocations, respectively.

There are some advantages of this geometric description of defects in solids. 
In contrast to the ordinary elasticity theory, this approach provides an 
adequate language for continuous distribution of defects. The problem of the 
description of quantum (or classical) dynamics of particles (or quasiparticles) 
in the elastic medium is reduced to a problem in a curved/torsioned space. In this 
framework, the influence of the defects on the  motion of electrons and 
phonons, for example, becomes reasonably easy to analyze, due to the fact that 
the boundary conditions imposed by the defects are incorporated into the 
geometry. In this geometric point view, the quasiparticles motion experience an 
effective non-Euclidean metric in an elastic media. This change in the 
effective metric experienced by the quasiparticles in this medium is caused by 
the stress and strain provoked in the elastic medium by topological defects. The 
classical motion of quasiparticles in the elastic background is described by 
geodesics in the effective metric.

In  this work we use the geometric theory of defects to describe a medium with 
disclinations. We are interested in the study of phonon  scattering in the 
presence of defects. In this context, phonons are represented by geodesics in 
the space that describes a deformed medium. Recently, the scattering of phonons 
by defects in solids was investigated making use of the geometric theory of defects~\cite
{Moraes96,dePaMo98,katanaev1}. In these previous analyses, the calculus of
geodesics was used to investigate the scattering of phonons in the presence of 
topological defects. In the present work we use  parallel  transport of
vetors around the defects to investigate the  Aharonov-Bohm effect 
in the scattering of phonons in the presence of a disclination and of multiple 
disclinations.

The geometric theory of defects shows the equivalence between three-
dimensional gravity with torsion and the theory of defects in solids. The 
defect acts as a source of a "gravitational" distortion field. The metric 
describing the medium surrounding the defect is then a solution to the three-
dimensional Einstein-Cartan equation. In a metric gravitational theory, a 
gravitational field is related to a nonvanishing Riemann curvature tensor. 
However, the presence of localized curvature can produce effects on the 
geodesic motion and also produce effects on parallel transport in regions where 
the curvature vanishes. A known example of this is provided when a particle is 
transported along a closed curve, which encircles an idealized cosmic 
string~\cite{vilenkin}. This situation corresponds to the gravitational 
analogue~\cite{ford} of the electromagnetic Aharonov-Bohm effect~\cite
{aharonov}. 

These effects are of  nonlocal origin and may be viewed as a manifestation 
of the nontrivial topology of the  spacetime of the cosmic string. It is worth 
calling  attention to the fact that, differently from the electromagnetic 
Aharonov-Bohm effect, which is essentially a quantum  effect, its gravitational 
analogue appears also at a purely classical level. Thus, in summary, the 
gravitational analogue of the electromagnetic Aharonov-Bohm effect, in this 
context, is the following: particles constrained to move in a region where the 
Riemann curvature tensor vanishes may exhibit a gravitational effect arising 
from a region of nonzero curvature from which they are excluded.

In the present case, the space that describes the defects is characterized by 
conical singularities~\cite{staro} in the curvature tensor. The conical 
singularity gives rising to curvature concentrated on the disclination 
axis. In this way, a medium that contains a disclination is described by nontrivial 
curvature concentrated on the defect axis. In this way, the quasiparticle 
 moving in the elastic medium with a disclination is moving in a region of 
null curvature. We investigate the existence of the Aharonov-Bohm effect in the 
movement of a quasiparticle in the presence of a disclination based in this 
physical observations. The solid state analog of the Aharonov-Bohm effect in 
semiconductors with dislocations has been investigated recently~\cite{tf}. In 
this case, it was shown that the occurrence of Aharonov-Bohm interference for 
particles moving around charged dislocations in a semiconductor gives {\it 
rise} to magnetoconductance oscillations in the macroscopic sample~\cite
{figielski}. Related to this, the existence of a geometrical phase in screw 
dislocations has been also investigated~\cite{ribeiro1}.

 Lately, Katanaev~\cite{katanaev2} has demonstrated that a geometric theory of 
defects is an equivalent description to nonlinear elasticity theory. He also 
showed that in the linear limit of his nonlinear geometric elasticity theory 
its solutions recovers the usual solutions of defects in a linear elasticity 
theory. Other alternative approaches to study this problem have been proposed 
which use either a gauge field~\cite{kadic,3,lasa} or a gravity-like 
approach~\cite{4,6}. Similar analogies hold in ($2+1$) dimensional gravity~\cite
{13,14}, where the spacetime geometry of a point particle can be understood in 
terms of distortions~\cite{puntigam}.

Our objective in this work is to demonstrate that an effect similar to the 
gravitational Aharonov-Bohm effect occurs in the scattering of phonons in an
elastic medium with disclinations. This letter is organized as follows, in the 
next section we consider the defects in an elastic medium. In the third 
section, we study the holonomy transformations associated to the deficit angle 
investigation in a space with disclinations. In the fourth section, we extend 
our analysis to a global characterization of multiple defects in a medium. 
In the last section, we present the concluding remarks.

\section{Defects in an Elastic Medium}

Defects can appear from thermal fluctuations, from extreme boundary conditions 
imposed or by the action of  external fields.A class of defects is called 
topologicaland is associated to some broken continuous symmetry and can be 
characterized by a core region where the order is destroyed and by an external 
region where the elastic properties are not changed. The physical systems can 
present broken continuous symmetry in which the elastic variables describe 
variations from a initial configuration. When we use a continuum approach we do 
not have information about microscopic details, however we can find a 
qualitative description. Usually, we use continuum models to represent a 
specific material. It could be based in information obtained by  
phenomenological constitutive models. We make an  idealization to describe an 
elastic solid where the action of the microscopic degrees of freedom is 
summed  up into few material parameters known as the elastic constants.

Recently, it was shown  a quantitative agreement between the standard 
elasticity theory and the geometric theory of 
defects~\cite{katanaev2}. The results  obtained by a linear approximation 
from the geometric theory agree with those of 
elasticity theory, showing that there is a good concordance between the 
treatments.

We consider a wedge dislocation as an infinite elastic medium without the z-
axis. This kind of defect, is constructed by 
a cut of an infinite wedge with the angle $-2\pi\theta$, which can be removed 
or inserted.  We start from the metric below 
which describes a wedge dislocation in the framework of the geometric theory of 
defects, 
\begin{equation}
\label{elastic}
ds^{2}=\left(\frac{r}{R}\right)^{(2\beta -1)}\left(dr^{2}+\frac{\alpha^{2}r^{2}}
{\beta^{2}}d\phi^{2}\right)+dz^{2},   
\end{equation}
with
\begin{equation}
\beta=\frac{-\theta\sigma+\sqrt{(\theta\sigma)^{2}+4(1+\theta)(1-\sigma)^{2}}}{2
(1-\sigma)},
\end{equation}
where $\sigma$ is the Poisson ratio and $\theta$ is an angle associated with 
the formation angle of the defect. This Poisson ratio is related with elastic 
properties in real materials, and is determined by the strain along the normal 
direction. It can be observed experimentally. Note that, if we use in the metric 
(\ref{elastic}) the following transformation
\begin{equation}
\xi=\frac{\alpha}{\beta}\left(\frac{1}{R}\right)^{\beta-1}r^{\beta}, 
\end{equation}
we obtain the conical metric  
\begin{equation}
ds^{2}=\frac{1}{\alpha^{2}}d\xi^{2}+\xi^{2}d\phi^{2} +dz^{2},
\end{equation}
that can also transformed by $\rho=\xi/\alpha$ into
\begin{equation}
ds^{2}=d\rho^{2}+\alpha^{2}\rho^{2}d\phi^{2} + dz^{2}.
\end{equation}
This  result is very interesting because it resembles the space 
of the cosmic string. The nonzero components of the Riemannian curvature tensor of the metric above is 
given by 
\begin{equation}
R^{12}_{12}=R^{1}_{1}=R^{2}_{2}=2\pi\frac{(1-\alpha)}{\alpha}\delta^{(2)}(\rho),
\end{equation}
which indicates that this metric corresponds to a locally flat medium with a 
conical singularity at the origin. 
Here, $\delta^{2}(\rho)$ is the two dimensional delta function in flat 
space. From the last expressions it 
follows that if $\alpha \in (0,1)$ the defect carries positive 
curvature and if $\alpha \in (1,\infty)$ the defect carries negative curvature.
 
The metric given by (\ref{elastic}) can be obtained by a linear
approximation in $\theta$, where $\beta \approx 1 +\frac{\theta(1-2\sigma)}{2(1-
\sigma)}$~\cite{katanaev2}.
\begin{eqnarray}
ds^{2}&=&\left(1+\theta\frac{1-2\sigma}{1-\sigma}\ln \frac{r}{R}\right)dr^{2}
\nonumber \\
&+&r^{2}\left(1+\theta\frac{1-2\sigma}{1-\sigma}\ln\frac{r}{R}+\theta\frac{1}{1-
\sigma}\right)d\phi^{2}.
\end{eqnarray}
This metric correspond the usual metric which is described by linear elasticity 
theory.

\section{Holonomy Transformation}

Our study of the analogue of gravitational Aharonov-Bohm  scattering of 
quasiparticles by a topological defect  is done by computing the 
holonomy matrix for some specific closed paths. When a vector is parallel 
propagated along a loop in a manifold $\textsl{M}$, the curvature of the 
manifold causes the vector, initially at $p\in \textsl{M}$, to appear rotated 
with respect to its initial orientation in tangent space $T_{p}\textsl{M}$, 
when  it returns to $p$. The holonomy is the path dependent linear 
transformation $T_{p}\textsl{M}{\rightarrow}T_{p}\textsl{M}$ responsible for 
this rotation. Positive and negative curvature manifolds, respectively, yield 
deficit or excess angles between initial and final vector orientation under 
parallel transport around such loops. This global property  of the manifold can 
be used as a means of global classification of spacetimes, as pointed out by 
Rothman, Ellis and Murugan ~\cite{cqg:rem}.

In recent works \cite{Moraes96,dePaMo98,katanaev1} the geodesics of particles 
in  the presence of topological defects has been calculated. In the eikonal 
approximation, or geometric optics, elastic deformations described by the wave 
equation
move along extremals. By analogy with motion of photons in electrodynamics it 
is natural to assume that extremals are
trajectories of phonons in elastic media with defects. In this way analysis of 
extremals yields a complete
picture of the scattering of phonons on disclinations. Here, the approach adopted 
by us consists in the utilization of the holonomy matrix to describe the phonon 
scattering in a disclinated medium. In this work we use the holonomy matrix to 
investigate in the classical level the analog of the Aharonov-Bohm effect. Essentially, 
holonomy is a matrix related to the notion of parallel transport of objects 
like vectors, spinors and tensors along curves in a given n-dimensional 
manifold. Associated to a  path $\gamma$, connecting points $A$ and $B$  in the 
manifold, the parallel transport matrix is given by the path-ordered exponential
\begin{equation}
\label{phase}
U(\gamma)={\cal P}\exp\left(
-\int_{A}^{B} \Gamma_{\mu}dx^{\mu}\right),
\end{equation}
where $\Gamma_{\mu}$ is the n-adic connection on the manifold and ${\cal P}$ is 
the path ordered product.
When the path $\gamma$ is closed, $U(\gamma)$ is known as the holonomy matrix. 
Notice the similarity of this equation with the Dirac phase factor $\Phi(\gamma)
={\cal P}\exp\left(ig\oint_{\gamma} A_{\mu}dx^{\mu}\right)$, the observable in 
the Aharonov-Bohm effect.

Holonomy has had an important role in the loop formulation of gauge 
theories~\cite{mandelstam1,wil}, of
quantum gravity~\cite{ash} and is a very convenient tool to obtain topological 
properties of specific 
geometries. For instance, Burges~\cite{prd:b} and Bezerra~\cite{valdir1} 
examined the effects of parallel 
transport of vectors and spinors both around a point-like solution and a 
cylindrically symmetric cosmic 
string pointing out to a gravitational analogue of the Aharonov-Bohm effect. 
More recently, holonomy has 
been used to study quantum computation in a geometric approach~\cite{ijmp:pz}.

The metric (\ref{elastic}) is described by a dual 1-form basis $e^{a}$, 
defined in terms of the 
dreinbeins $e^{a}_{\mu}$ by: $e^{a}=e^{a}_{\mu}dx^{\mu}$, where
\begin{subequations}
\label{elastic-basis}
\begin{eqnarray}
        e^{1}&=& dz    \mbox{,}  \\
        e^{2}&=& \left(\frac{r}{R}\right)^{(\beta-1)}dr  \mbox{,}   \\
        e^{3}&=& \left(\frac{r}{R}\right)^{(\beta-1)}\frac{\alpha r}{\beta}
d\phi  \mbox{.}   \\
\end{eqnarray}
\end{subequations}
In this way the metric is written as $ds^{2}=(e^{1})^{2}+(e^{2})^{2}+(e^{3})^{2}
$. We need to determine the 1-form connection $\omega^{a}_{b}$ in order to find 
the holonomy transformations for the "space-time" of a defect. The 1-form 
connection satisfies Cartan's structure equation, $T^{a}=de^{a}+\omega^{a}_
{b}\wedge e^{b}$,  where $T^{a}$ is the torsion 2-form. Using the fact that the 
correspondent geometry is torsion-free, we find the following 1-form connections
\begin{eqnarray}
\omega^{3}_{2}&=&-\omega^{2}_{3}=\alpha d\phi    \mbox{,} 
\end{eqnarray}
the other terms are null. The connection 1-forms transform in the same way as 
the gauge potential of a non-Abelian gauge theory. The above connections leads 
to the following spin connection
\begin{eqnarray}
\Gamma_{\phi}=
\left(
\begin{array}{ccc}
0 & 0 & 0   \\
0 & 0 & -\alpha  \\
0 & \alpha & 0  
\end{array}
\right).
\end{eqnarray}
Note that this matrix ${\Gamma}_{\phi}$ can be written in terms of the 
generator of rotation $J_{12}$ about the z-axis
\begin{equation}
\label{eq4}
\Gamma_{\phi}= i \alpha J_{12}.
\end{equation}

Let us analyze the parallel transport of vectors around closed curves that 
contains the singularity. 
In our case the unique contribution to the holonomy is provided by the 
azimuthal spin connection. Thus, we may write:
\begin{equation}
U(\gamma_{\phi})={\cal P}\exp \left(-\oint \Gamma_{\phi}d\phi \right).
\end{equation}
Making the expansion of this expression and noticing that we are always able to 
write the exponents of upper order 
in the $\Gamma_{\phi}$ and $\Gamma_{\phi}^{2}$ terms, the above equation can be 
written in the following way
\begin{eqnarray}
U(\gamma_{\phi})=1-\frac{\Gamma_{\phi}}{\alpha}\sin(2\pi\alpha)+ \left(\frac
{\Gamma_{\phi}}{\alpha}\right)^{2}\left[1-\cos(2\pi\alpha) \right].
\end{eqnarray}
Alternatively one can express the holonomy in a matrix form
\begin{eqnarray}
U(\gamma_{\phi})=
\left(
\begin{array}{cccc}
1 & 0 & 0 &  \\
0 & \cos (2\pi \alpha) &\sin (2\pi \alpha)  \\
0 &-\sin (2\pi\alpha) &\cos (2\pi \alpha)
\end{array}
\right).
\end{eqnarray}
Considering the linear approximation of the term $\beta$ we can rewrite the 
holonomy matrix in terms of the elastic parameters
\begin{eqnarray}
U(\gamma_{\phi})=
\left(
\begin{array}{cccc}
1 & 0 & 0 &  \\
0 & \cos \left[2\pi\frac{\beta(2-\sigma)-1}{(1-2\sigma)}\right] & \sin \left[2
\pi\frac{\beta(2-\sigma)-1}{(1-2\sigma)}\right]   \\
0 & -\sin \left[2\pi\frac{\beta(2-\sigma)-1}{(1-2\sigma)}\right] &\cos \left[2
\pi\frac{\beta(2-\sigma)-1}{(1-2\sigma)}\right]
\end{array}
\right),
\end{eqnarray}
where
\begin{equation}
\beta \approx 1+\theta\frac{(1-2\sigma)}{2(1-\sigma)}.
\end{equation}
The holonomy group measures the deviation of the space from global flatness.

The deficit angle $\chi$ is obtained when we compare final and initial 
positions of the parallel transported 
vector and it is given by
\begin{equation}
\cos \chi_{a}=U_{a}^{a},
\end{equation}
where $a$ is the tetradic index. The terms of non-vanishing angular deviations 
occurs when $a=1$ and $a=2$, so we have
\begin{equation}
|\chi|=|2\pi\alpha+2\pi j|,
\end{equation} 
where $j$ is an integer. When $\alpha \rightarrow 0$ we must have $\chi=0$, so 
we choose $j=0$ which leads to
\begin{eqnarray}
|\chi|&=&|2\pi\alpha|\nonumber\\
			&=&\left|2\pi\frac{\beta(2-\sigma)-1}{(1-2\sigma)}
\right|.
\end{eqnarray} 
The above expression shows that $\chi\neq0$, if we parallel transport a vector 
around a closed path the final vector does not coincides with the original 
vector. This physical effect could be understood as a gravitational analogue of 
the Aharonov-Bohm effect which is denominated elastic Aharonov-Bohm effect.
\section{Global characterization of the multiple Defects}
In this section we use the previous holonomy matrix to investigate the Aharonov-
Bohm effect for phonon scattering,in a medium with $N$ parallel disclinations. 
We use the global characterization process to obtain the phase acquired by the
vector
when it is parallel transported in a closed path around the defects~\cite
{valdir:cqg}. As an application of the holonomy transformation we study the 
space configuration from the global point of view of $N$ defects located at 
points $a_{j}, j=1, 2, \dots, N $. In order to do this we use the result of the 
previous section that only defects
enclosed by the circle contribute to the phase factor acquired by a vector  
parallel transported along the circle, in the background  of the multiple 
defects. If we parallel transport a vector $\vec{X}$ around a circle enclosing 
a chiral magnetic string we get the following vector after this process:
 \begin{eqnarray}  
 \vec{X}^{(1)}= U_{1}\vec{X} ,\label{b26}    
  \end{eqnarray}   
 \noindent where $U_{1}$ is obtained from
\begin{eqnarray}
U_{k} =\exp[-2i \pi \alpha_{k} J_{12}]  , \label{b27}
\end{eqnarray}
\noindent
taking $k$ equal to one.

Now, let us consider a system of two defects, the defect 1 at$\vec{a}_{1}=0$ 
(origin) and defect 2 at $\vec{a}_{2}$. If we take a vector $\vec{X}$ and carry 
it along a circle around defect $2$, the resulting vector is given by $U_{2}\vec
{X}$.
We then carry this resulting vector parallel to a circle around defect $1$. 
Then, we get the following result
\begin{eqnarray}
\vec{X}^{(2)}=\vec{b}_{1,2} + U_{1}U_{2}\vec{X},  \label{b28}
\end{eqnarray}
\noindent
where $\vec{b}_{1,2}=U_{1}(1-U_{2})\vec{a}_{2}$.

If we consider a system of three defects, we have
\begin{eqnarray}
\vec{X}^{(3)}=\vec{b}_{1,2,3} + U_{1}U_{2}U_{3}\vec{X},  \label{b29}
\end{eqnarray}
\noindent
where $\vec{b}_{1,2,3}=U_{1}(1-U_{2})\vec{a}_{2} +
U_{1}U_{2}(1-U_{3})\vec{a}_{3}.$.
It is easy to generalize this result for a system of N defects, located, 
respectively,  at $\vec{a}_{1}, \vec{a}_{2}, \dots, \vec{a}_{N}$. The vector 
$\vec{X}^{(N)}$ obtained after the parallel transport of a vector $\vec{X}$ is 
given by the expression
\begin{eqnarray}
\vec{X}^{(N)}=\vec{b}_{1,2, \dots ,N} + U_{1}U_{2}\dots
U_{N}\vec{X},\label{b30}
\end{eqnarray}
\noindent
where $\vec{b}_{1,2, \dots ,N}=U_{1}(1-U_{2})\vec{a}_{2} +U_{1}U_{2}(1-U_{3})
\vec{a}_{3} + \dots + U_{1}U_{2}\dots
U_{N-2}(1-U_{N})\vec{a}_{N-1}U_{N-1}(1-U_{N})\vec{a}_{N}.$, and $U_{N}$ is 
given by Eq.(\ref{b27}) with $k=N$. Then, a vector $\vec{X}$ parallel 
transported in the field of N defects acquires a phase factor given by $U_{1}U_
{2}\dots U_{N}$ and from the global point of view, the system of defects 
behaves like a simple string with matching conditions given
Eq.(\ref{b30}).

Let us consider a single defect that behaves like this system. So, if we parallel
transport a vector $\vec{X}$  around a circle, in 
whose center we have an equivalent defect  of deficit angle characterized by 
$\alpha _{g}$, we get the following vector after this process
\begin{eqnarray}
\vec{X}^{N+1} =  U\vec{X}
\label{c1}
\end{eqnarray}
\noindent
where $\vec{a}$ is the position of defect string that is equivalent to the 
system of defects.

Equating the geometrical phases acquired by a vector $\vec{X}$ in both cases, 
we have
\begin{eqnarray}
U_{1} U_{2} \cdots U_{N-1} U_{N}=  U . \label{c2}
\end{eqnarray}
Taking the trace of Eq.(\ref{c2}) we obtain
\begin{eqnarray}\label{cos}
  \cos\phi = \cos(\sum_{j=1}^{N-1}\phi_{j})\cos\phi_{N}
 -\sin(\sum_{j=1}^{N-1}
\phi_{j})\sin\phi_{N} 
\end{eqnarray}
\noindent
where $\phi_{j}= 2\pi\alpha_{j}$ and $\phi= 2\pi\alpha $. Using some 
trigonometric manipulation we can write the expression (\ref{cos}) in the 
following form
\begin{eqnarray}\label{cos1}
  \cos\phi = \cos(\sum_{j=1}^{N}\phi_{j})
\end{eqnarray}  
This is a relation between the deficit angles of the resulting space and the 
deficit angle of the defect involved. We have thus demonstrated that, in the 
global point of view, the  $N$ topological defects can be seen as an equivalent 
defect with defect angle given by
\begin{equation}
\phi=\sum_{j=1}^{N}\phi_{j}.
\end{equation}
Since the space outside the region of multiple defects is locally flat, we can describe 
the analytic solution purely in terms of
space patches with Euclidean metric, but connected by some matching conditions 
which are given by
\begin{equation}
\left(\begin{array}{c}
z'\\
x'\\
y'
\end{array}\right)=\left(\begin{array}{ccc}
1&0&0\\
0&\cos(\phi)& \sin(\phi)\\
0&-\sin(\phi)& \cos(\phi)\\
\end{array}\right) \left(\begin{array}{c}
z\\
x\\
y
\end{array}\right), \label{b19}
\end{equation}
\noindent
that relates points $(\vec{X})$ and ${(\vec{X'})}$ along the  edges. Equation 
(\ref{b19}) is the exact expression for the holonomy, for circles in the space 
of multiple defects. Only the defect surrounded by the circle 
contributes to the change in this vector.

The existence of locally flat coordinates in this space permits us to consider 
Eq.(\ref{b19}) as a parallel transport matrix. So we can say that when a vector 
is carried out along a circle in this space-time it acquires a phase that 
depends on $\alpha_{i}$, which prevents it from being equal to the unit matrix. 
This effect is due to the non-trivial topology of the space under 
consideration. This is the gravitational analogue Aharonov-Bohm effect for $N$ 
linear defects in solids.
\section{Concluding Remarks}
We have shown that the geometric theory of defects can be 
applied in order to investigate classical effects due to influence of the 
topology. We have analyzed in this work the scattering of phonons in a 
disclinated media using  loop variables, which allowed us to obtain various physical 
and geometric properties in a specific background. We have considered a 
geometric point view in order to describe the motion of quasiparticles  experiencing 
 an effective non-Euclidean metric in an elastic media associated with the 
presence of disclinations. 

We have known that a single dislocation produces a dilation that breaks the 
translational symmetry of the crystaline lattice, producing a deformation of 
the geodesics around the defect. We have demonstrated that a single disclination 
modifies the medium, and provokes a break of rotational 
symmetry which is noted in the vector parallel transport framework.

In our final result, which is given by Eq.(\ref{b19}), we notice that the 
change in the phase of the quantum wave function for a particle, incorporates the 
effect of the effective deficit angle associated with the topological defects. 
We also studied the analogue of the gravitational Aharonov-Bohm effect for 
scattering of quasiparticles by topological defects. This  was done by computing 
the holonomy matrix for some specific closed paths. We have used the matrix of 
parallel transport to study the Aharonov-Bohm effect in this background. We 
have demonstrated that an effect similar to the gravitational Aharonov-Bohm effect 
occurs for quasiparticles  moving in this effective metric. We have 
analyzed this effect in an elastic medium with one and with $N$ defects via global 
characterization. In this way, we have demonstrated that the scattering of 
phonons by one or more disclinations produces a global effect in an elastic 
medium, that we denominated elastic Aharonov-Bohm effect, due to deformations 
caused by defects in this medium.
\section*{Acknowledgment}
We thank  CAPES (PROCAD), CNPq, PRONEX and Funda\c{c}\~ao de Amparo \`a 
Pesquisa do Estado da Bahia(FAPESB) - Brazilian agencies - for financial supports.
We thank Professor F. Moraes for the critical reading of this manuscript.

\end{document}